\documentclass[10pt]{article}
\usepackage{amsmath}
\usepackage{epsfig}
\newcommand{\eq}{\begin{equation}}
\newcommand{\qe}{\end{equation}}
\newcommand{\eqr}{\begin{eqnarray}}
\newcommand{\rqe}{\end{eqnarray}}

\newcommand{\U}{u_{\infty}} 
\begin{document}
\begin{flushright}
Feb 11, 2003 \\
\end{flushright}
\vspace{10mm}

\begin{center}

{\Large{\bf Another Leigh-Strassler deformation through 
the Matrix model}}
\\
\vspace{10 mm}
Teresia M{\aa}nsson  \footnote{email: teresia@physto.se}
\vspace{4mm}\\
Institute of Theoretical Physics,\\ Box 6730, SE-113 85 Stockholm, Sweden

\vspace{15mm}

{\bf Abstract }
\end{center}

\noindent
In here the matrix model approach, by Dijkgraaf and Vafa, is used in 
order to obtain the effective superpotential for a certain deformation
 of N=4 SYM discovered by Leigh and Strassler. 
An exact solution to the matrix model Lagrangian is found and is expressed 
in terms of elliptic functions.

%\newpage
\setcounter{page}{1}

\section{Introduction}
Recently Dijkgraaf and Vafa  proposed a method using the matrix model to 
calculate the effective superpotential for N=1 supersymmetric gauge
theories  \cite{DijkgraafVafa1},  \cite{DijkgraafVafa2}, \cite{DijkgraafVafa},
\cite{Cachazo}.
 In particular they showed for the mass
 perturbation of N=4
SYM theory down to an N=1 theory, that the effective superpotential
 agreed with the result by Dorey \cite{dorey:elliptic} obtained with
 different methods.

One class of interesting N=1 superconformal theories are the ones
discovered by Leigh and Strassler as marginal and relevant
 deformations of N=4 SYM.
 We use the matrix model to extract the effective superpotential
for a particular relevant perturbation of N=4 SYM, discovered
by Leigh and Strassler \cite{leigh:strassler}. 
 This 
superpotential is very similar to the one studied by 
Dorey et.al. \cite{dorey:Leigh}, it differs though through the fields in
the q-deformed commutator, in their case they are $\Phi^+=\Phi_1+i\Phi_2$
 and $\Phi^-=\Phi_1-i\Phi_2$ and in our case we have $\Phi_1$ and 
$\Phi_2$ therefore it could be of interest to really understand the
physical differences between them. In the case when the deformed 
commutator becomes an ordinary commutator they both reduce to the
mass deformed theory mentioned above.

Even though this case looks at first sight a bit more difficult
than the case studied by Dorey etc, it has some nice properties
that will be seen. For instance it transforms in a nice way under 
S transformation and the eigenvalue distribution is symmetrical distributed
around zero. We expect that these nice properties are due to the symmetries
between exchange of the fields which is absent in the other \cite{dorey:Leigh}
 case.
% that is 
%described with the potential
%\eq
%W=Tr\left(\Phi_1(e^{i\beta}\Phi_2\Phi_3-e^{-i\beta}\Phi_3\Phi_2)
%-m\sum \Phi_i^2 \right)
%\qe
%The $\Phi$'s are chiral fields.
%There have been a lot studying on the model when you have that 
%$\Phi_2=\Phi_3^{\dagger}$. In the case 

\section{The action and the elliptic world}
The superpotential of the Leigh-Strassler deformed SYM that will be 
investigated looks like
\begin{equation}
W=- Tr \left( e^{i\beta}\Phi_1\Phi_2\Phi_3-e^{-i\beta}\Phi_1\Phi_3\Phi_2)-
m\sum \Phi_i^2 \right)
\end{equation}
where the fields $\Phi_i$ are chiral transforming under $U(N)$.
Now we would like to apply the method of Dijkgraaf and Vafa in order to 
find an effective superpotential of the $U(N)$ theory in its confining
vacua. Accordingly we should look at the
partition function
\begin{equation}
Z=\int \mathcal{D}\Phi 
e^{\frac1{g_s} Tr \left( e^{i\beta}\Phi_1\Phi_2\Phi_3-e^{-i\beta}\Phi_1\Phi_3\Phi_2)-
m\sum \Phi_i^2 \right)}
\end{equation}
and expand it around the classical vacuum where $\Phi_{cl}=0$.
 The fields are all to be treated as
hermitian matrices.
 It would be nice to see that even in 
this case the equations of motions for the quantum fluctuation will yield
an elliptic structure, and indeed we will see that that is the case.
Integrating out $\Phi_2$ and $\Phi_3$ and diagonalizing $\Phi_1$
the partition function looks like
\eq
\int \prod d\lambda_i \prod_{j<i}
\frac{(\lambda_i-\lambda_j)^2}
{(\lambda_i+\lambda_j)^2\sin^2\beta+(\lambda_i-\lambda_j)^2\cos^2\beta+4m^2}
\prod_i \frac{e^{-\frac{m}{g_s}\lambda_i^2}}{\sqrt{\lambda_i^2\sin^2\beta+m^2}}
\qe
Here we see that the effective action look like something with an external
quadratic potential well and the numerator as a coulomb force for electrons
moving in two dimensions, the denominator looks a bit more tricky with
both a repulsive part and an attractive. The classical minimum will be in 
the minimum for the potential well, which is $\lambda=0$. There is no force
pulling it more in the negative than the positive direction, so in the
quantum regime the eigenvalues will be distributed symmetrically around
$\lambda=0$.
Making the following change in coordinates 
\eq
\lambda_i=2m\frac{\sinh \mu_i}{\sin 2\beta}
\qe
the partition function transforms into
\eqr
\int \prod d\mu_i \prod_{j<i}
\frac{(\sinh \mu_i-\sinh \mu_j)^2}
{(\sinh \mu_i-\sinh (\mu_j+2i\beta)(\sinh \mu_i-\sinh (\mu_j-2i\beta)}\\
\prod_i \frac{e^{-4m^3\sinh^2 \mu_i/g_s\sin^2 2\beta}\cosh \mu_i}
{\sqrt{\sinh \mu_i-\sinh (\mu_i+2i\beta)}
\sqrt{\sinh \mu_i-\sinh (\mu_i-2i\beta)}}
\rqe
a bit of algebra makes it look like up to some constant:
\eqr
\label{langrangian}
\int \prod d\mu_i \prod_{j<i}
\frac{(\sinh (\frac{\mu_i-\mu_j}{2})\cosh(\frac{\mu_i+\mu_j}{2}) )^2}
{\sinh (\frac{\mu_i-\mu_j}{2}-i\beta)\cosh(\frac{\mu_i+\mu_j}{2}+i\beta)
\sinh (\frac{\mu_i-\mu_j}{2}+i\beta)\cosh(\frac{\mu_i+\mu_j}{2}-i\beta)}\\
\prod_i \frac{e^{-4m^3\sinh^2 \mu_i/g_s\sin^2 2\beta}\cosh \mu_i}
{\sqrt{\cosh (\mu_i+i\beta)}
\sqrt{\cosh (\mu_i-i\beta)}}
\rqe
Now we rescale the $\mu_i$'s with a factor of one half,
 because it will be convenient later on
and calculate the equations of motions from the effective action
\eqr
\label{EOM}
\frac{2m^3}{g_s\sin^2 2\beta}\sinh x-\frac1{2}\tanh \frac{x}{2}+
\frac1{4}\tanh (\frac{x}{2}+i\beta)+\frac1{4}\tanh (\frac{x}{2}-i\beta)=\\
\sum_{j}\left[\frac1{2}\coth(\frac{x-\mu_j}{4})-
\frac1{4}\coth(\frac{x-\mu_j}{4}+i\beta)
-\frac1{4}\coth(\frac{x-\mu_j}{4}-i\beta)\right]+\\
\sum_{j}\left[\frac1{2}\tanh(\frac{x+\mu_j}{4})-
\frac1{4}\tanh(\frac{x+\mu_j}{4}+i\beta)-
\frac1{4}\tanh(\frac{x+\mu_j}{4}-i\beta)\right]
\rqe
Here $x$ is one of the eigenvalues for which we will solve the 
equation of motion.
We will use the fact that the
eigenvalues $\lambda_i$ are distributed symmetrical around $\lambda=0$
then it implies that also $\mu_i$ will be distributed symmetrical
 around $\mu=0$, thus they will take values between say $-a$ and $a$.
A resolvent can then be  introduced (see \cite{kostov})
\eq
w(z)=\frac1{2}\int dx \frac{\rho(x)}{\tanh(\frac{z-x}{2})}
\qe
where $\rho(x)=\frac1{N}\sum \delta(x-\mu_i)$ and $w(-x)=-w(x)$.
The resolvent will have a cut along the eigenvalues, $z\in [-a, a]$, 
and the jump in the resolvent along the cut will give the density
\eq
-2\pi i\rho(x)=w(x+i\epsilon)-w(x-i\epsilon)
\qe
and the normalisation condition on the density can be expressed like
\eq
1=\int_{-a}^a dx \rho (x)=\frac1{2\pi i}\oint_C w(x)
\qe
where $C$ is a curve going around the cut.
 Then the equations of motion
 can be written like
\eqr
\frac{2m^2}{\sin^2 2\beta}\sinh x-\frac{S}{2N}\tanh \frac{x}{2}+
\frac{S}{4N}\tanh \left(\frac{x+2 i \beta}{2}\right)
+\frac{S}{4N}\tanh \left(\frac{x- 2 i \beta}{2}\right) =\\ \frac{S}{N}
\sum_{j}\left[\coth(\frac{x-\mu_j}{2})-
\frac1{2}\coth\left(\frac{x-\mu_j+4i\beta}{2}\right)
-\frac1{2}\coth\left(\frac{x-\mu_j-4i\beta}{2}\right)\right]
\rqe
and the right hand side can be written in terms of the resolvent
\eq
S\left(2w(x)-w(x+4i\beta)-w(x-4i\beta)\right)
\qe
here we introduced the 't Hooft coupling $S=g_s N$.
 This can in turn be written  as
\eq
\label{bla}
g(x+2i\beta)-g(x-2i\beta)
\qe
where
\eq
g(x)=w(x-2i\beta)-w(x+2i\beta)
\qe
notice that $g(x)$ is periodic when $x$ is shifted with $ 2\pi i$.
We would also like to rewrite the left hand side in the same manner, 
with the same constant.
Let us have a look at the following function
\eq
\label{residual}
h(x)=-i\xi\cosh x +
\frac{S}{4N}\sum_{k=1}^n\left[\tanh \left(\frac{x+k 4 i \beta}{2}\right)
-\tanh \left(\frac{x- k 4 i \beta}{2}\right)\right]
\qe
where $\xi=m^3/ \sin^3 2\beta$. We see that
\eq
h(x+2i\beta)-h(x-2i\beta)
\qe
will be the same as the left side of (\ref{EOM}) if $\beta=\pi l/(2n+1)$, where
$l$ is any integer such that $\beta$ is in between zero and $\pi/2$.
Here it should be mentioned that we have an arbitrariness in our
choice of $h(x)$, it is defined up to a constant,
 but in the end that constant would have to
be removed.
Then in that case the equations of motion can be expressed like
\eq
\label{blabla}
J(x+2i\beta)-J(x-2i\beta)=0, \;\;\;\;\; x\in [-a,a]
\qe
where $x$ is in between $a$ and $-a$ and where $J(x)=g(x)-h(x)$, to
write it more explicit 
\eq
\label{J(z)}
J(x)=i\xi\cosh x -
\frac{S}{4N}\sum_{k=1}^n\left[\tanh \left(\frac{x+k 4 i \beta}{2}\right)
-\tanh \left(\frac{x- k 4 i \beta}{2}\right)\right]+S(w(x-2i\beta)-w(x+2i\beta))
\qe
A small comment before proceeding, the Lagrangian from which 
the equation of motion was derived only contained a potential term
so that the force on a probe eigenvalue 
$\lambda_i=x$ is equal to the derivative of the potential with
respect to $x$ which is nothing other than
\eq
\label{force}
f(x)=-J(x+2i\beta)+J(x-2i\beta).
\qe
Keep this in mind because it will be of use when the superpotential
is derived.
We can analytical continue $J(x)$ into the complex plane, and it is 
clear that
$J(z)$ is periodic with period $2\pi i$ and in the strip $|Im z|<\pi$,
it is holomorphic besides the two cuts at $Re z \in [-a,a]$ and
 $Im z=\pm 2i\beta$ and  simple poles at $z_k=4ik\beta-i\pi+i 2\pi j$,
 where $k$ is
an integer between $1$ and $n$ and the integer $j$ is chosen such that $z_k$ 
is in the interval $-\pi$ and $\pi$. Another way to write $z_k$ which might be
more illuminating is 
\eq
\label{zedkey}
z_k=\pi\frac{2k-1}{2n+1}(-1)^{[\frac{k-1}{m}]}(-1)^{n+1}
\qe
where the bracket $[\,]$ stands for the integer part of what is inside.
In particular if we glue everything together we get a torus with poles in it
and it should be possible to express it then as an elliptic function.
An elliptic function is determined by its poles and the asymptotic
behavior.
\begin{figure}
\center
\epsfig{figure=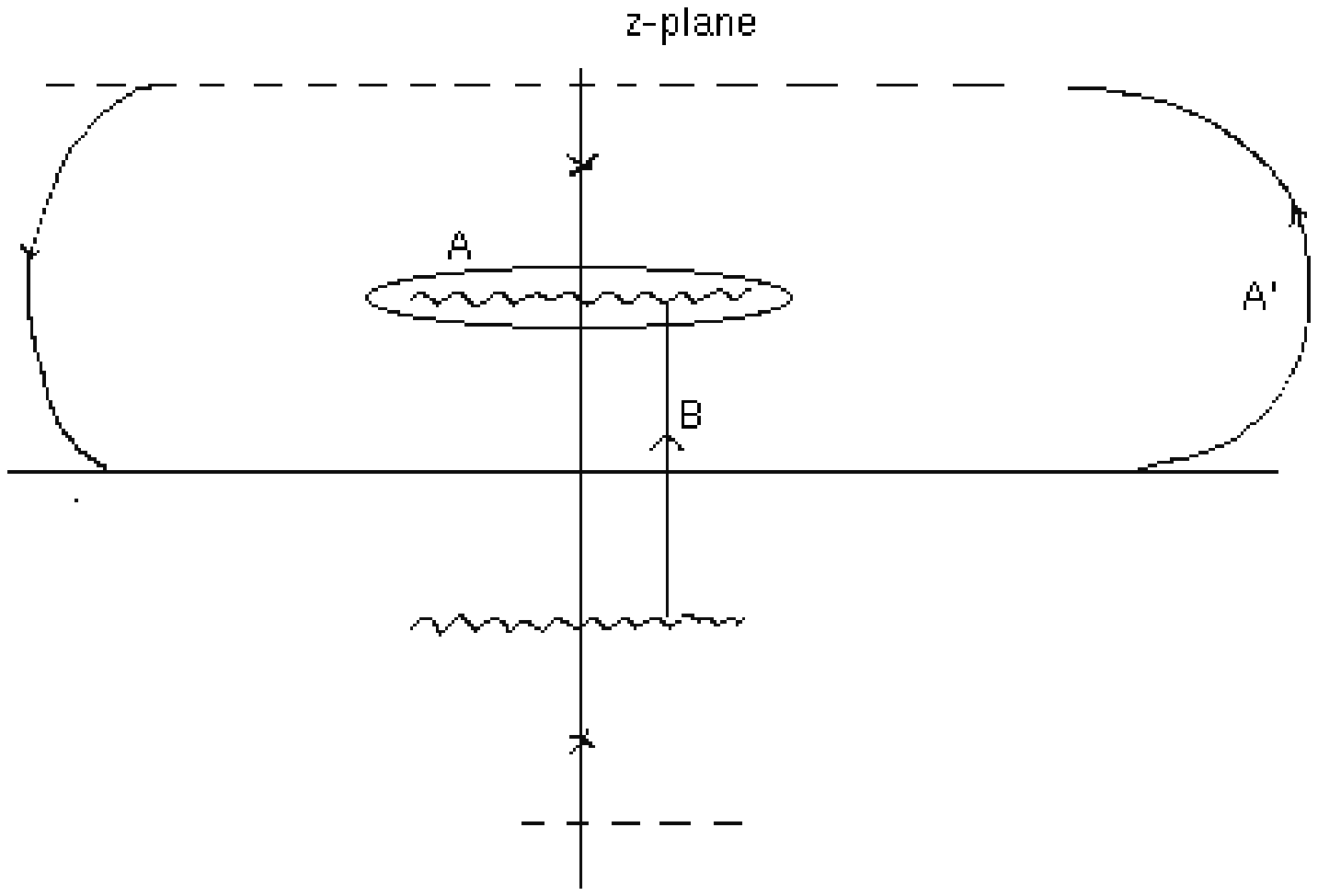, height=6.7cm}
\epsfig{figure=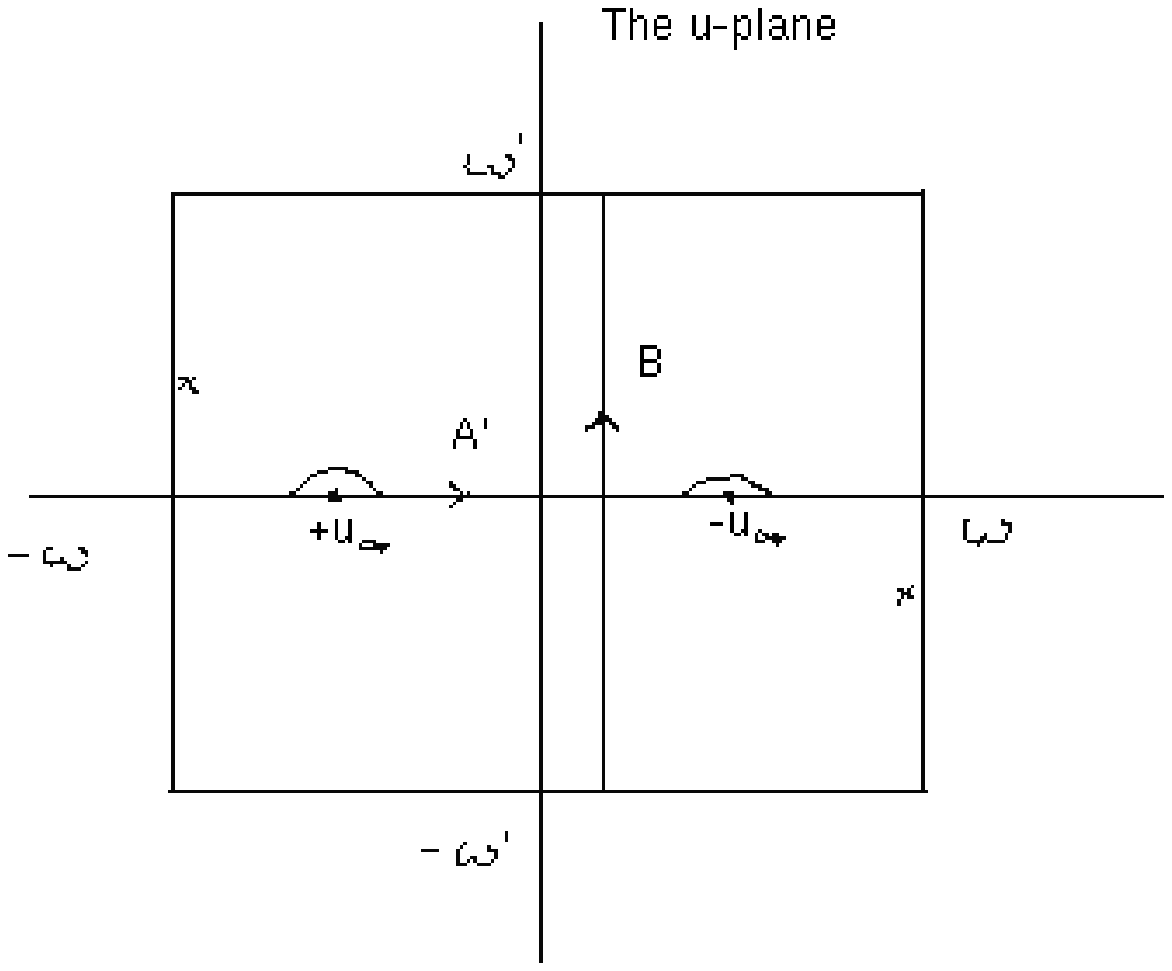, height=6.7cm}
\caption{Here you can see the z-plane and the u-plane and how
the loops are mapped. The imaginary axis between the cuts in the z-plane
are mapped to the imaginary axis in the u-plane, where $\omega=0$
and the imaginary axis above the cut in the z-plane is mapped to
the line $u=\pm \omega$ and $z=\pm 2\beta$ goes to $u=\pm \omega'$.}
\label{figur:kontur}
\end{figure} 
The asymptotic behavior we can see from $h(z)$ both when we let
$z\rightarrow \pm\infty$. $h$ is symmetric in $z$.
\eq
\label{infinity}
J=i\frac{\xi}{2} e^{\pm z}
\qe
The structure of our equations are very much the same as
in \cite{kostov} and we could use similar techniques in order
 to solve the problem, 
the main differences is that here $J$ is symmetric around $z=0$ and
instead of a double pole at plus infinity, we have one simple pole
at plus infinity and one at minus infinity and also several other
poles coming pairwise.  $J$ is a doubly periodic function, 
one period coming from going around one cut and the other from going 
from one cut to the other, thus
 we can parametrize our system with 
$u$ such that 
\eqr
\label{period}
z(u+2\omega)=z(u), & z(u+2\omega')=z(u)+4i\beta \\
J(u+2\omega)=J(u), & J(u+2\omega')=J(u)
\rqe
From \cite{kostov} we get a  regular map between $z$ and $u$
that satisfies the conditions (\ref{period})
\eq
\label{zed}
\exp{z(u)}=\frac{H(u_\infty+u)}{H(u_\infty-u)}=
\frac{\theta_1(\pi u/2\omega-\beta+\pi/2)}
{\theta_1(\pi u/2\omega+\beta+\pi/2)}
\qe
if $\U=(\pi-2\beta) K/\pi$ where $K$ is the standard quarter period and
 coincide with $\omega$
(see figure (\ref{figur:kontur}) to see the $z$-plane and corresponding
$u$-plane). Sometimes we will use the notation $\beta'=\beta-\pi/2$,
because it is practical when dealing with the $\theta_1$ functions.
We use the following Ansatz for the Elliptic function $J$
\eq
\label{ws:formal}
J(u)=A+\frac{B}{\wp(u)-\wp(\U)}+\sum_{u_k}\frac{C_k}{\wp(u)-\wp(u_k)}
\qe
where $u_k$ is the value of $u$ that corresponds to the poles 
$z_k $, thus this elliptic function has simple poles
 at $u=\pm \U$ and $u=\pm u_k$,
there are a lot of equivalent ways of doing this ansatz.
The constants $C_k$ can be determined through the residues at
$\pm z_k$ in equation (\ref{residual}) and
 the constants $A$ and $B$ can be determined by  an expansion around
$u_\infty$. First expand the expression (\ref{ws:formal}) and then do 
the same for the original $J$ expressed in terms of the theta functions.
%If we use equation bla bla for $J$ and put in bla bla we get the
%behaviour at infinity as
%From here we get A and B expressed in terms of $H$ and $sn$.
See Appendix for further details, there  the following expressions
for the constants are obtained
\eq
B=-i\frac{\xi}{2}\wp' (\U)\frac{H(2\U)}{H'(0)}
\qe
\eq
A=\frac{B\wp''(\U)}{\wp'^2(\U)}-\frac{C}{\wp(\U)-\wp(u_k)}+
i\frac{\xi}{2}
\frac{H'(2u_\infty)}{2H'(0)}
\qe
\eq
C=\frac{S}{2N}\wp'(u_k)
\left(z'(u)
\right)^{-1}.
\qe
%%%%%%%%%%%%%%%%%%%%%%%%%%%%%%%%%%%
%%%%%%%%%%%%%%%%%%%%%%%%%%%%%%%%%%%%%%%%%%%%
\subsection{The Superpotential}
Now we would like to apply the methods of Dijkgraaf and Vafa to
obtain the effective superpotential. The effective superpotential
has the following form:
\eq
W_{eff}(S)=N\frac{\partial F_0}{\partial S}-2\pi\,i \tau_0 S,
\;\;\; \delta W_{eff}(S)=0
\qe
where $\tau_0=\theta/2\pi+i4\pi/g^2_{YM}$ is the gauge coupling, 
 physics is invariant under a change of $\theta\rightarrow \theta +2\pi$.
 This also has to be 
extremized  in order to get the effective potential, for the
Leigh-Strassler deformation, as a function of
the coupling constant $\tau_0$.
 The great thing is 
that these terms in the expression for the effective superpotential
 can be expressed in terms of loop integrals
over the loops $A$ and $B$, see figure (\ref{figur:kontur}).
From the expression for $J(z)$ (see (\ref{J(z)}) it is clear  that S  can be expressed
in terms of  a line integral around one
of the cuts of $J(z)$.
\eq
2\pi i S=\Pi_A=\int_{C_A} J(z)dz=\int_{-\omega+\omega'}^{\omega+\omega'}
 J(u)z'(u)du
\qe
Here we have to be a bit careful because there are also the singularities
at $\pm z_k$, (see figure (\ref{figur:kontur}) and equations (\ref{zedkey}) and
(\ref{residual})), with residues 
 $S/2N$ up to a sign, thus integrating around
 the $A'$
loop instead leads to
\eq
2\pi i S \left(1+\frac{ n'}{N}\right)=
\Pi_A'=\int_{C_{A'}} J(z)dz=\int_{-\omega}^{\omega}
 J(u)z'(u)du
\qe
where
\eq
n'= \sum(-1)^{[\frac{k-1}{l}]}=\pm \frac1{2}\mbox{rem}(\frac{2l-n}{2l}),
\qe
there is a positive sign in front if $n$ is odd and a negative sign 
if $n$ is even, and $2 n$ is the numbers of  poles, (this is
the same $n$ which appeared earlier e.g. in the expression for $\beta$). The 
contribution $n'$ you get from the poles in the upper half $z$-plane.
Observe that for $l=1$ and $n$ even, $n'$ is zero.

The other part $\partial_S F_0$ is the derivative of the planar
free energy and 
 can be calculated from integrating the force on the probe eigenvalue
(\ref{force}) from the infinity to the eigenvalue cut, but also
here we should be careful, just as in the other Leigh-Strassler 
deformation studied by Dorey et.al., there is also a ``zero point''
contribution to the free energy.
 From considering
the Lagrangian (\ref{langrangian}), in the exponent there was the
 term $2\sinh^2(x/2)$
this could have been rewritten like $(\cosh x -1)$, that is we get 
a zero point energy contribution from that constant, which more
explicitly looks like $-2 N m^3/g_s\sin^2 2\beta$, the zero genus free
energy is this times $g_s^2$, that is $-2Sm^3/\sin^2 2\beta$. Thus the
total derivative of the free energy with respect to $S$ becomes
\eq
 \frac{\partial F_0}{\partial S}=\int_{\infty}^a f(x)-
2\frac{m^3}{\sin^2 2\beta}.
\qe
 The first part
can be rewritten as an integral over $J(z)$ going from the end of the
 upper cut to infinity and then back from infinity to the end of the
lower cut,  and can in turn be transformed to a
line integral  going directly from  the upper
cut to the lower. The singularities in the $z$-plane is all
placed along the imaginary axis thus there are no problems deforming
the integral.
\eq
 \int_{\infty}^a f(x)=\Pi_B=\int_{C_2}
 J(z)dz=\int_{-\omega'}^{\omega'} J(u)z'(u)du
\qe
One thing you could have been worried about concerning this deformation 
is  closing the loop  at infinity. In the $u$ coordinates 
 the integrand above, $J(u)z'(u)$, can be expanded
around the point in $u$ corresponding to the infinity, there it will 
consist of a double pole and thus the residue will be zero, thus there is
no problem in closing at infinity for the $J(z)$, which we constructed.
An extra constant in $J$ would have to be removed in order for this 
to work.

 We see that the effective superpotential
expressed in terms of the loop integrals:
\eq
W_{eff}(S)=N\Pi_B-\frac{\tau_0\Pi_{A'}}{1+\frac{ n'}{N}}
-2\frac{N m^3}{sin^2 2\beta}.
\qe
Then extremizing the superperpotential
\eq
\delta W_{eff}=N\delta \Pi_B-\frac{\tau_0\delta\Pi_{A'}}{1+\frac{ n'}{N}}
=0
\qe
The loop integrals have some nice properties and the loop integral
around $B$ can be expressed as (see Appendix)
\eq
\Pi_B=\left( \tau+\frac{1}{\pi}\frac{\sum v_k}{N- n'} \right)\Pi_A
+2\xi \frac{\theta(2\beta)}{\theta'(0)}
\qe 
and
\eq
\delta \Pi_B=\left( \tau+ \frac{1}{\pi}
\frac{\sum v_k}{N- n'} \right)\delta \Pi_A
\qe
here we introduced the notation $v_k=\pi u_k/2\omega$.
This gives us the relation between the  gauge coupling and the
elliptic parameter $\tau$
\eq
\label{tau}
 \tau\left(N+n'\right)+\frac{1}{\pi}
\sum v_k =\tau_0
\qe
and the effective superpotentials values at the different $\tau$s 
\eq
\label{superpotential}
W_{eff}=
2N\xi\frac{\theta(2\beta|\tau)}{\theta'(0|\tau)}-2N\xi\sin 2\beta .
\qe
Here $\tau$ should satisfy the solution above (\ref{tau}), but because of the
non-trivial $\tau$ dependence which $v_k$ has we will look at two
different limits.
The summation over the $v_k$ will be a constant when $\tau$ is not too  
small or/and when $\beta$ goes to zero, which can be seen to be
\eq
z_k=\ln{\frac{\theta_1(\beta'-v_k)}{\theta_1(\beta'+v_k)}}=
\ln{\frac{\sin(\beta'-v_k)}{\sin(\beta'+v_k)}}
+\sum_1^{\infty}\frac1{n}\frac{q^{2n}}{1-q^{2n}}\sin 2nv_k\sin 2n\beta'
\qe
q will be very small for most $\tau$'s and then the last sum will 
be negligible. Two cases can occur, either the $v_k$ are pure imaginary
or $v_k=i\alpha\pm\pi/2$.
 Then the relation between $v_k$ and $z_k$ can be expressed like
\eq
\tan v_k=\cot\beta \tanh z_k/2
\qe
In this case $\tau$ becomes
\eq
\tau=\frac1{N-n'}(\tau_0-j-i\alpha)
\qe
where $j$ is an integer or a half integer, if it is an integer it will
 correspond to a shift by $\theta$ with $2\pi$
 and $\alpha$ just the sum over all the imaginary 
values in $u_k$ thus the effect is that $\tau_0$ gets shifted by a constant and
also $N$ get shifted.   
In the limit
$\beta$ goes to zero in such a way that $n$ will be an even number,
$\tau$ will become $\tau=\tau_0/N$ and  equation (\ref{superpotential})
 will coincide with the mass perturbed case
calculated by Dijkgraaf, Vafa \cite{DijkgraafVafa} and Dorey 
\cite{dorey:elliptic}.

How $v_k$ behaves for small values of $\tau$ can be seen from using the modular
properties of the $\theta_1$ functions
\eqr
z_k=\ln{\frac{\theta_1(\beta'-v_k|\tau)}
{\theta_1(\beta'+v_k|\tau)}}=
\ln{\frac{e^{-i(v_k-\beta')^2/\pi\tau}}{e^{-i(v_k+\beta')^2/\pi\tau}}
\frac{\theta_1((\beta'-v_k)/\tau|-1/\tau)}
{\theta_1((\beta'+v_k)/\tau|-1/\tau)}}\\
\approx 4i v_k \beta'/\pi\tau+
\ln{\frac{\sin((\beta'-v_k)/\tau)}{\sin((\beta+v_k)/\tau)}}
%=4i u_k \beta/\pi\tau+1/\tau
\rqe
We have to look at two different cases, the case when $v_k$ is purely
imaginary, then the second term becomes $2i v_k/\tau$ and in the 
other case the second term becomes $2i\beta/\tau$. In the first case
then 
\eq
4i v_k \approx \tau \pi \frac{z_k}{\beta}=\tau\pi 
\frac{(2k-1)}{l}(-1)^{[\frac{k-1}{m}]}(-1)^{n+1}\equiv\tau\pi L_k
%=\pi\tau \left[\frac{km-n'}{m-n'} \right]
\qe
And the second case we have
\eq
\begin{split}
4i v_k &=\frac{\pi\tau }{\beta-\pi/2}(z_k-i\pi(-1)^p)-2\pi i \\
 &=2\pi\tau\left[\frac{2k-1-(2n+1)}{2l-2n-1} \right]
(-1)^{[\frac{k-1}{m}]}(-1)^{n+1}-2\pi i(-1)^p \equiv
\pi\frac{\tau}{2}M_k-2\pi i (-1)^p
\end{split}
\qe
where $p$ is an even number if $z_k$ is positive and odd if $z_k$ is 
negative.
For the expression of $z_k$ see equation (\ref{zedkey}).
The following relation holds between $\tau$ and $\tau_0$
\eq
 \tau =
(\tau_0+j)/\left(N- n'+\sum_k{L_k/4}+
\sum_k{M_k/2}) \right)
\qe
The second summation is over the $k$ which fulfill the condition
\eq
|2k-1|>2l
\qe
and the first summations over the other $k$'s.
So we see that the effect is that $\tau_0$ get shifted with an half
 integer or an integer
and $N$ gets shifted with something that sometimes is an integer and
for some special values of $\beta$ it could be zero.

To get the gluino condensate we differentiate the effective superpotential
(\ref{superpotential}) with respect
to $\tau_0$, that will also correspond to the $S$ we calculated in
the appendix. Remember though to see that you have to remember that
the relation between $\tau_0$ and $\tau$ is in general a bit complicated.
Anyway taking the mass decoupling limit, letting $m\rightarrow \infty$
together with $\tau \rightarrow i\infty$ such that $m^3 q^2=\Lambda^3$,
where $\Lambda$ is a constant, we see that we get that the effective
superpotential becomes a constant independent of $\beta$ as it should. 
%%%%%%%%%%%%%%%%%%%%%%%%%%%%5 
%%%%%%%%%%%%%%%%%%%%%%%%%%%%%%
\section{Classical vacua, confinement and S-duality}
In \cite{dorey:Nstar}  they found that the
underlying S-duality from the $N=4$ SYM was realized, in the
mass perturbed  case, via modular transformation on the gauge coupling
 relating the superpotential for
different massive vacua and a similar thing was found in \cite{dorey:Leigh}.
In \cite{dorey:Leigh} they extract an  $SU(N)$ version of the
Leigh-Strassler deformation they looked at and  found for instance
that the confining vacuum was related to the higgs vacuum  via S-duality. 
Here we will look at the eigenvalue distribution around the classical 
$\Phi_{cl}=0$ and see what will happened to it if we perform an 
S-transformation and then take its classical limit. We will see 
that it is proportional to some of the classical vacua
of the theory.

Berenstein et.al. \cite{berenstein} looked at the classical solutions to
the same model as we are considering. That is they started from the
superpotential
\eq
 W= Tr \left( \phi_1\phi_2\phi_3-q\phi_1\phi_3\phi_2)+
m\sum \phi_i^2 \right).
\qe
The F-flatness condition gave them the following relation between the
fields
\eq
[\phi_j, \phi_{j+1}]_q=\phi_{j+2}, \;\;\;\mbox{cyclic on $j$, mod $3$}
\qe
from which they derived some of the vacua of the theory.
 The $q$-commutator is defined
through $[a,b]_q=ab-qba$. They did a field redefinition in order to 
get rid of the $m$, in the same fashion a field redefinition can be made 
in order to get the commutator in the form we have. Let
 $\phi\rightarrow q^{1/2}\phi$, then the commutator becomes
\eq
 q^{-1/2}\phi_i\phi_j-q^{1/2}\phi_j\phi_i.
\qe
Anyway in order to get the vacua they had to find irreducible
 representations to
this algebra. A certain class of these representations will be
 deformed $sl(2,C)$ representations, thus in the case $q=1$ the
algebra takes the form of $sl(2,C)$.
 First they noted that there exist a one dimensional representation
for $q\neq 1$ looking like
\eq
\phi_j=\frac1{1-q}
\qe
and a two dimensional irreducible representation looking like
\eq
\phi_j=\frac{-i}{1+q}\sigma_j
\qe
where the $\sigma_j$ are the Pauli matrices. This latter one can be looked 
upon as a deformation of $sl(2,C)$ representation. Then they came up with an 
ansatz to find a general form of a deformed $sl(2,C)$ representation.
In the case of even dimension $2p$
 of the representation they found the following eigenvalues
to the matrices $\phi_i$
\eq
\pm\alpha_n=\pm\frac1{q^{p-n}(q^{-1/2}+q^{1/2})}\sigma_{2(p-n)}[q]
\qe
where  $\sigma_x[q]=1+q+q^2+\ldots +q^x$, here
we have accounted for the rescaling of the field $\phi$.
In the case $q=e^{2i\beta}$   this can in
 fact be rewritten as
\eq
\label{classicaleigen}
\pm\alpha_n=\pm \frac{\sin(\beta (2(p-n)-1))}{\sin 2\beta}
\qe
And in the case of odd dimension $2p+1$ of the representation, the eigenvalues 
they obtained look like
\eq
0, \;\pm\alpha_n=\pm\frac{\sigma_{(p-n)}[q^2]}{q^{p-n}}
\qe
They warn  that these solutions are not D-flat, but
the solutions is related to a D-flat solution via an $SL(M)$
transformation.

In \cite{dorey:Leigh} they conjecture a method for getting the values for
the eigenvalues of the condensate of the field $\Phi$, we will see 
that this method indeed gives us the right type of expectation value
 of $\Phi^2$, and performing an S-transformation on
those eigenvalues  reproduces something proportional to
 the vacua above
in the classical limit $\tau\rightarrow i\infty$.
In order to get the eigenvalue distribution the idea is now first to 
notice that the eigenvalues $\lambda_i$ to the field $\Phi_1$ is
\eq
\lambda_i=2m\frac{\sinh \frac{\mu_i}{2}}{\sin 2\beta}
\qe
where $\mu_i$ is between $-a$ and $a$. The cut in the resolvent that
emerges in $J$ is displaced from the position  of the
eigenvalues $\mu_i$ with a distance $i 2\beta $ if we are considering 
the upper cut in the $z$ plane for $J$. Their proposal is then to
displace  $\mu_i$ with a distance $i2\beta$ and then evaluate
the function $\lambda(x)$ where the $i$'s have been made continuous along 
the upper cut (the A-cycle)
\eqr
\lambda (x)=2m\frac{\sinh \frac{( x-2i\beta)}{2}}{\sin 2\beta}
=\frac{m}{\sin 2\beta}\left(
e^{i\beta}\frac{\theta_1^{1/2}(\pi x+\tau \pi/2-\beta)}
{\theta_1^{1/2}(\pi x+\tau \pi/2+\beta)}
-e^{-i\beta}\frac{\theta_1^{1/2}(\pi x+\tau \pi/2+\beta)}
{\theta_1^{1/2}(\pi x+\tau\pi/2-\beta)}
\right)
\\
=\frac{m}{\sin 2\beta}\left(
\frac{\theta_4^{1/2}(\pi x-\beta|\tau_0/N')}
{\theta_4^{1/2}(\pi x+\beta|\tau_0/N')}-
\frac{\theta_4^{1/2}(\pi x+\beta|\tau_0/N')}
{\theta_4^{1/2}(\pi x-\beta|\tau_0/N')}
\right) \;\;\;\; x \in \left[0, 1\right]
\rqe
here $N'$ stands for the shifted $N$ we got when solving the equations
of motions. We will not be so careful about the shift and just considering 
$N'$ as if it were the usual $N$, a more careful treatment should bee 
done, but now we  look at things very roughly.
It is easy to check in the classical limit $\tau_0\rightarrow i\infty$ the 
expression above goes to zero, which is expected because we looked
at quantum fluctuation around the classical eigenvalue zero. 
Then they claim that the eigenvalues should be uniformly distributed
along this and the expectation value of the field squared should be given
by integrating  the square of this over the range of $x$ gives.
\eq
\langle \lambda^2(x) \rangle=N\int_0^1 \lambda(x)=
\frac{2Nm^2}{\sin^3 2\beta}
\frac{\theta (2\beta)}{\theta'(0)}-\frac{2Nm^2}{\sin^2 2\beta}
\qe
Again the last part is just due to the zero energy. Here we see that
this at least are at agreement with the expression of the
effective superpotential (\ref{superpotential}), to differentiate the
superpotential with respect to the mass should give as the
expectation value of the field $\Phi^2$.

Now we will 
see which nice properties this $\lambda(x)$ has under S-transformation, 
letting $\tau_0\rightarrow -1/\tau_0$ the eigenvalues $\lambda(x)$
becomes
\eq
\lambda (x)
=\frac{m}{\sin 2\beta}\left(
\frac{\theta_3^{1/2}(\pi x-\beta|-1/\tau_0N')}
{\theta_3^{1/2}(\pi x+\beta|-1/\tau_0N')}
-\frac{\theta_3^{1/2}(\pi x+\beta|-1/\tau_0 N')}
{\theta_3^{1/2}(\pi x-\beta|-1/\tau_0N')}
\right) \;\;\;\; x \in \left[-\frac1{2}, \frac1{2}\right]
\qe
Now we like to see what will happened with this in the classical 
limit $\tau\rightarrow i\infty$. It is easy to take the limit if 
we use the modular transformation rules of the theta function
\eq
\theta_3(x|\tau)=
(-i\tau)^{1/2}e^{-i x^2/\pi \tau}\theta_3(x/\tau|-1/\tau)
\qe
using this we get that the $\lambda(x)$ equals
\eq
\lambda (x)
=\frac{m}{\sin 2\beta}\left(
e^{-2i\beta x \tau N}\frac{\theta_3^{1/2}(\tau_0 N(\pi x-\beta)|\tau_0N)}
{\theta_3^{1/2}(\tau_0 N(\pi x+\beta)|\tau_0N)}
-e^{2i\beta x \tau N}\frac{\theta_3^{1/2}(\tau_0 N(\pi x+\beta)|\tau_0 N)}
{\theta_3^{1/2}(\tau_0 N(\pi x-\beta)|\tau_0N)}
\right)
\qe
The theta three's goes to one in the limit $\tau\rightarrow i\infty$.
Thus after doing an S-transformation on $\lambda(x)$ wee see that it 
becomes in the classical limit
\eq
\lambda(x)=im\frac{\sin (2\beta x \tau N)}{\sin 2\beta}
\qe
We could say that also $\beta$ are supposed to transform under S  like
$\beta\rightarrow \beta/\tau_0$ \cite{dorey:Leigh} then the above expression
 can be
written like
\eq
\lambda(x)=im\frac{\sin 2\beta}{\sin 2\beta/\tau}
\frac{\sin (2\beta x  N)}{\sin 2\beta}
\qe
This expression is up to the strange factor in front of it
equal to the classical eigenvalues (\ref{classicaleigen}) if
you make a discretization of $x$.  
%%%%%%%%%%%%%%%%%%%%%%%%%%%%%%%%%%5
%%%%%%%%%%%%%%%%%%%%%%%%%%%%%%%%%%%%%
\section{Conclusions}
First of all an exact solution to the Leigh-Strassler deformation
under consideration was found and was shown to have of an elliptic
structure. The solution was found for special values of the parameter
$\beta$, which parametrize the deformation, these $\beta$'s were the ones
that could be written as some fraction of $\pi$. 
 The solution was used to apply 
the method proposed by Dijkgraaf and Vafa, to find an effective
superpotential for the Leigh-Strassler deformation and thus the
gluino condensate.
It looks very similar to the effective potential of the other Leigh-Strassler 
deformation which has been studied in this contexts, but there are some 
crucial differences as the symmetry around the expectation value of the
field $\Phi$ and its elliptic parameter $\tau$ has a more complicated
 dependence on the gauge coupling constant
 $\tau_0$, which might be of interest. For some special $\beta$
you get the same simple $\tau_0$ dependence as in the earlier studied cases.

Then we had a brief look at how $S$-transformation acts on the 
solution and saw that the confining vacuum was related 
to some higgs vaccum in the classical limit.
%This could be an indication that the $S$-duality from the $N=4$ theory
%could be realized in this model just as in \cite{dorey:Nstar} and 
%\cite{dorey:Leigh}, but it is worth further studies.
%%%%%%%%%%%%%%%%%%%%%%%%
%%%%%%%%%%%%%%%%%%%%%%
\begin{appendix}
\section{Calculations}
For all the relations between elliptic functions and theta functions
and the integrals over them see \cite{byrd:elliptic} and 
\cite{as:handbook}
\subsection{The coefficients}
The constants $C_k$ can be determined by the fact that the residue of
 (\ref{residual}) around $u_k$ should coincide with the residual of
(\ref{ws:formal}) around $u_k$.
Do a Laurent expansion of $\tanh$ around $x=z_k$ in 
(\ref{residual})
\eq
\frac{S}{4N}\tanh\left(\frac{x-z_k}{2}\right)=
\frac{S}{2N}\frac1{x-z_k}
\qe
Then using $x=z(u)$, where $z(u)$ is given by \ref{zed} and expanding
 it around $u_k$ and
put it in the equation above
\eq
\frac{S}{2N}\frac1{z'(u)(u-u_k)}
\qe
Expanding (\ref{ws:formal}) around $u_k$ gives
\eq
\frac{C}{\wp'(u_k)}\frac1{u-u_k}
\qe
And from putting the residuals equal C becomes
\eq
C=\frac{S}{2N}\wp'(u_k)\left( z'(u)
\right)^{-1}
\qe
Now we can decide the constants $A$ and $B$ through doing an expansion around
$u_\infty$. First do it for the expression (\ref{ws:formal})
\eq
J(u)=A+\frac{B}{\wp'(\U)(u-u_\infty)}-\frac{B\wp''(\U)}{2\wp'^2(\U)}
+\frac{C}{\wp(\U)-\wp(u_k)}
\qe
If we use equation (\ref{infinity}) for $J$ and put in the expression
for $z(u)$ in terms of theta functions we get the
behaviour at infinity as
\eq
J(u)=i\frac{\xi}{2}\left(\frac{H(2u_\infty)}{H'(0)(u_\infty-u)}
-\frac{H'(2u_\infty)}{H'(0)}\right)
\qe
From here we get A and B expressed in terms of theta and elliptic functions
\eq
B=-i\frac{\xi\wp' (\U)}{2}\frac{H(2\U)}{H'(0)}
\qe
\eq
A=\frac{B\wp''(\U)}{\wp'^2(\U)}-\frac{C}{\wp(\U)-\wp(u_k)}-
i\xi
\frac{H'(2u_\infty)}{H'(0)}
\qe
%%%%%%%%%%%%%%%%%%%%%%%%%%%%%%%%%%%%%
\subsection{The integral calculations}
The two integrals of interest in order to get the effective superpotential
is:
\eq
\Pi_{A'}=\int_{C_{A'}} J(z)dz=\int_{-\omega}^{\omega}
 J(u)z'(u)du
\qe
And
\eq
\Pi_B=\int_{C_B} J(z)dz=\int_{-\omega'}^{\omega'} J(u)z'(u)du
\qe
\eqr
z'(u)=\frac{H'(u+\U)}{H(u+\U)}-
\frac{H'(u-\U)}{H(u-\U)}=
\zeta(u+\U)-\zeta(u-\U)-2\zeta(\omega_1)\frac{\U}{\omega} =\\
-\frac{\wp'(\U)}{\wp(u)-\wp(\U)}+2\zeta(\U)-2\zeta(\omega_1)\frac{\U}{\omega}
=-\frac{\wp'(\U)}{\wp(u)-\wp(\U)}+
\frac{H'(2\U)}{H(2\U)}-\frac{\wp''}{2\wp'}(\U)
\rqe
The integral can be divided in the following pieces,
 which can be rewritten in standard
elliptic integral forms. First the piece consisting of the poles $\U$ having
the constant $B$ in it
\eqr
\Pi_1=-B\wp' (\U)\int \frac1{(\wp(u)-\wp(\U))^2}+
B \left(\frac{H'(2\U)}{H(2\U)}-\frac{\wp''}{2\wp'}\right)
\int \frac1{\wp(u)-\wp(\U)}  =\\
-B\left[ \frac{\wp''(\U)}{\wp'^2}\ln{\frac{\sigma(u+\U)}{\sigma(u-\U)}}
-\frac1{\wp'(\U)}(\zeta(u+\U)+\zeta(u-\U))\right. \\
\left.-\left(\frac{2\wp(\U)}{\wp'(\U)}+ 
\frac{2\wp''(\U)\zeta(\U)}{\wp'^2(\U)}\right)u +
\frac1{\wp'}\left(\frac{\wp''}{2\wp'}-
\frac{H'(2\U)}{H(2\U)}\right) 
\left(\ln{\frac{\sigma(u-\U)}{\sigma(u+\U)}}
+2u\zeta(\U)\right)
\right]
\rqe
and the part consisting of $C$
\eqr
\Pi_2=C\left(\frac{-\wp'(\U)}{\wp(u_k)-\wp(\U)}+
\frac{H'(\U)}{H(\U)} \right) 
\int \frac1{\wp(u)-\wp(u_k)}\\
+\frac{1}{\wp(u_k)-\wp(\U)}\int\frac{C\wp'(\U)}{\wp(u)-\wp(\U)} =\\
-\frac{S}{2N}\left[ \ln{\frac{\sigma(u-u_k)}{\sigma(u+u_k)}}
+2u\zeta(u_k)\right]+\frac{C}{\wp(u_k)-\wp(\U)}
\left[ \ln{\frac{\sigma(u-\U)}{\sigma(u+\U)}}
-2u\zeta(\U)\right]
\rqe
The only contribution of the integral over the constant $A$ comes from
 the $C_2$ curve
\eqr
\Pi_3=A\int_{C_2} dz=4i\beta\left(
 \frac{B\wp''(\U)}{2\wp'^2(\U)}-\frac{C}{\wp(\U)-\wp(u_k)}-
i\xi
\frac{H'(2u_\infty)}{2H'(0)}\right)
\rqe
The terms in $\Pi_A$ and $\Pi_B$ linear in $u$ will just relate
the integrals with a factor of $\tau=\omega'/\omega$,
 that is $\Pi_B=\tau\Pi_A$ 
(linear in $u$). 
From looking at the integrands we see that we need to know
\eqr
\left[\ln{\frac{\sigma(u+\U)}{\sigma(u-\U)}}\right]_{-\omega}^{\omega}=
\ln{\frac{\sigma(\omega+\U)}{\sigma(\omega-\U)}}-
\ln{\frac{\sigma(-\omega+\U)}{\sigma(-\omega-\U)}}\\
2\ln{\frac{\sigma(\omega+\U)}{\sigma(\omega-\U)}}=
2\ln{\frac{\theta_1(\pi/2-\beta+\pi/2)}{\theta_1(\pi/2+\beta-\pi/2)}}
+4\eta\beta/\pi
=4\eta\beta/\pi
\rqe
and
\eqr
\left[\ln{\frac{\sigma(u+\U)}{\sigma(u-\U)}}\right]_{-\omega'}^{\omega'}=
2\ln{\frac{\theta_1(\tau\pi/2-\beta+\pi/2)}{\theta_1(\tau\pi/2+\beta-\pi/2)}}+
4\tau\eta\beta/\pi
=4i\beta+4\tau\eta\beta/\pi
\rqe
And last we need
\eqr
\left[\zeta(u+\U)+\zeta(u-\U)\right]_{-\omega}^{\omega}=4\eta\\
\left[\zeta(u+\U)+\zeta(u-\U)\right]_{-\omega'}^{\omega'}=4\eta', \;\;\;
\; \eta'=\tau\eta-\pi/2\omega_1
\rqe
Now we see that
\eq
\Pi_B=\tau\Pi_A'+2\xi\frac{\theta(2\beta)}{\theta'(0)}+\frac{S}{2N}\sum 4i v_k
\qe
Using $S=N\Pi_A/((N-n')2\pi i)$ the above becomes
\eq
\Pi_B=\tau\left( 1+\frac{1}{\pi}\frac{\sum v_k}{N-n'}\right)\Pi_A'+2\xi\frac{\theta(2\beta)}{\theta'(0)}
\qe
\subsection{Useful relations between elliptic and theta functions}
\eq
\zeta(u)-\frac{\zeta(\omega_1)u}{\omega_1}=\frac{\pi}{2\omega_1}
\frac{\theta_1'(\pi u/2\omega_1)}{\theta_1(\pi u/2\omega_1)}
\qe
\eq
\ln{\frac{\theta(\alpha-\beta)}{\theta(\alpha+\beta)}}=
\ln{\frac{\sin(\alpha-\beta)}{\sin(\alpha+\beta)}}
+\sum_1^{\infty}\frac1{n}\frac{q^{2n}}{1-q^{2n}}\sin 2n\alpha\sin 2n\beta
\qe
\eq
2\eta(\alpha)=\eta(2\alpha)-\frac{\wp''(\alpha)}{\wp'(\alpha)}
\qe
\eq
\sigma(z)=\frac{2\omega}{\pi}\exp{\frac{\eta z^2}{2\omega}}
\frac{\theta_i(v)}{\theta'(0)}
\qe
\end{appendix}
\subsection*{Acknowledgments:}
I would like to thank Ansar Fayaduzzin for useful discussions and Gabriele
Ferretti for reading the manuscript.

\end{document}